# Metastable Bound States of the Two-Dimensional Bi-magnetoexcitons in the Lowest Landau Levels Approximation

S.A. Moskalenko[1,*], P.I. Khadzhi[1], I.V. Podlesny[1], E.V. Dumanov[1],

I.A. Zubac[1] and M.A. Liberman[2]

[1]Institute of Applied Physics, Academic str. 5, MD-2028 Chisinau, Republic of Moldova

[2]Nordic Institute for Theoretical Physics (NORDITA) KTH and Stockholm University, Roslagstullsbacken 23, SE-106 91 Stockholm, Sweden

*e-mail: exciton@phys.asm.md

**Abstract**

The possible existence of the bound states of the interacting two-dimensional (2D) magnetoexcitons in the lowest Landau levels (LLLs) approximation was investigated using the Landau gauge description. The magnetoexcitons taking part in the formation of the bound state with resultant wave vector $\vec{k}=0$ have opposite in-plane wave vectors $\vec{k}$ and $-\vec{k}$ and look as two electric dipoles with the arms oriented in-plane perpendicularly to the corresponding wave vectors. The bound state of two antiparallel dipoles moving with equal probability in any direction of the plane with equal but antiparallel wave vectors is characterized by the variational wave function of the relative motion $\varphi_n(\vec{k})$ depending on the modulus $|\vec{k}|$. The spins of two electrons and the effective spins of two holes forming the bound states were combined separately in the symmetric or in the antisymmetric forms $(\uparrow\downarrow + \eta \downarrow\uparrow)$ with the same parameter $\eta = \pm 1$ for electrons and holes. In the case of the variational wave function $\varphi_2(k) = (8\alpha^3)^{1/2} k^2 l_0^2 \exp[-\alpha k^2 l_0^2]$ the maximum density of the magnetoexcitons in the momentum space representation is concentrated on the in-plane ring with the radius $k_r = 1/(\sqrt{\alpha} l_0)$. The stable bound states of the bimagnetoexciton molecule do not exist for both spin orientations. Instead of them, a deep metastable bound state with the activation barrier comparable with the ionization potential $I_l$ of the magnetoexciton with $\vec{k}=0$ was revealed in the case $\eta=1$ and $\alpha=0.5$. In the case $\eta=-1$ and $\alpha=3.4$ only a shallow metastable bound state can appear.

# 1. INTRODUCTION

As was established earlier [1-3] the two-dimensional magnetoexcitons with wave vectors $\vec{k} = 0$ in the lowest Landau levels (LLLs) approximation do not interact at all between them due to their hidden symmetry, and are forming an ideal Bose gas. The hidden symmetry is related with the radii of the Landau quantization orbits, which do not depend on the effective masses of the electrons and holes and are determined exclusively by the magnetic length $l_0$.

In the Landau gauge description if the electron and the hole forming the exciton are moving in opposite directions with $q = -p$ they are deviated in the same gyration point $pl_0^2$ and undergo quantum oscillations around of it, what gives rise to the Landau quantization (LQ). The orbits of the quantum oscillations are the same for the electron and for the hole being determined by the magnetic length and such magnetoexciton with electron and hole on the lowest Landau levels and with center-of-mass wave vector $\vec{k} = 0$ look as completely neutral compound particles up till the excited Landau levels or the Rashba spin-orbit coupling are not taken into account, as it was done in Ref. [4].

The magnetoexcitons with center-of-mass wave vector $\vec{k}$ different from zero have the structures of the electric dipoles as is represented in Fig. 2.

The arms of the dipoles are oriented in-plane perpendicularly to the wave vectors $\vec{k}$:

$$\vec{\rho}_0 = \left[\vec{k} \times \vec{z}\right] l_0^2; \quad l_0^2 = \frac{\hbar c}{eB}. \tag{1}$$

Here $\vec{z}$ is the unit vector perpendicular to the layer and $B$ is the magnetic field strength. Due to the in-plane spatial separation of the electron and hole orbits, two magnetoexcitons with wave vectors different from zero do interact. This interaction can be described as a dipole-dipole interaction only in the case, when the middle distance between them is much greater that the electric dipoles arms.

The strong dependence between the relative electron-hole and the center-of-mass motions in the frame of the spin waves as well as in the case of the magnetoexcitons was underlined in the Ref. [5-7]. The difference between these two processes consists in the fact that the spin waves appear in the conditions of the completely filled lowest Landau levels, whereas the magnetoexcitons are described in the conditions of almost empty Landau levels.

It was shown that two spin waves moving in-plane in the same direction attract one another which leads to their binding. Two spin waves moving in opposite directions undergo repulsion [7]. Olivares-Robles and Ulloa [8] have investigated another model of bilayer magnetoexcitons with the electrons in conduction band of one layer and with the holes in the valence band of another spatially separated parallel layer of the double quantum well (DQW) subjected to the action of a strong perpendicular magnetic field. Electrons and holes undergo Landau quantization in different parallel planes and are bound by their Coulomb interaction forming the magnetoexcitons, which can move in-plane direction. Such magnetoexcitons are characterized by two types of dipole moments. One type is formed by static dipole moments $\vec{p}_\perp$ perpendicularly oriented to the planes of the layers. Another type arises due to the in-plane motion of the magnetoexcitons. This type of the dipole moments $\vec{p}_\parallel$ are oriented in-plane. The total dipole moment of such magnetoexcitons $\vec{p}$ consists from two components $\vec{p} = \vec{p}_\perp + \vec{p}_\parallel$. The perpendicular dipole moments being taken alone give rise to the repulsion between the magnetoexcitons with wave vectors $\vec{k} = 0$. In the Ref. [8] was shown that the total interaction between two moving magnetoexcitons with different from zero wave vectors $\vec{k}_1$ and $\vec{k}_2$ is more repulsive for their antiparallel orientations ($\vec{k}_2 = -\vec{k}_1$) and is less repulsive in the case of parallel orientations ($\vec{k}_1 = \vec{k}_2$). These conclusions are in full agreement with the results of the Ref. [7]. The paper is organized as follows. In section 2 the Hamiltonian, the spin structure and the wave functions in the lowest Landau levels (LLLs) approximation are presented. In section 3 the main obtained results are discussed. The section 4 contains the conclusions.

## 2. THE HAMILTONIAN OF THE ELECTRON-HOLE SYSTEM AND THE SPIN STRUCTURE OF THE BIMAGNETOEXCITONS

In the Ref. [9] the Hamiltonian describing the Coulomb interaction of the 2D electrons and holes situated on the lowest Landau levels (LLLs), taking into account the spins of the conduction electrons and the effective spins of the holes in the valence band was determined. For simplicity the electron-hole exchange Coulomb interaction was not taken into account. The Hamiltonian has the form

$$H_{Coul}^{LLL} = H_{e-e}^{LLL} + H_{h-h}^{LLL} + H_{e-h}^{LLL},$$

$$H_{e-e}^{LLL} = \frac{1}{2} \sum_{\vec{Q}} \sum_{p,q} \sum_{\sigma_1,\sigma_2} W(\vec{Q}) e^{-iQ_x Q_y l_0^2} e^{iQ_y(p-q)l_0^2} a_{p,\sigma_1}^+ a_{q,\sigma_2}^+ a_{q+Q_x,\sigma_2} a_{p-Q_x,\sigma_1},$$

$$H_{h-h}^{LLL} = \frac{1}{2} \sum_{\vec{Q}} \sum_{p,q} \sum_{\sigma_1,\sigma_2} W(\vec{Q}) e^{iQ_x Q_y l_0^2} e^{-iQ_y(p-q)l_0^2} b_{p,\sigma_1}^+ b_{q,\sigma_2}^+ b_{q+Q_x,\sigma_2} b_{p-Q_x,\sigma_1},$$

$$H_{e-h}^{LLL} = -\sum_{\vec{Q}} \sum_{p,q} \sum_{\sigma_1,\sigma_2} W(\vec{Q}) e^{iQ_y(p+q)l_0^2} a_{p,\sigma_1}^+ b_{q,\sigma_2}^+ b_{q+Q_x,\sigma_2} a_{p-Q_x,\sigma_1}, \quad W(\vec{Q}) = \frac{2\pi e^2}{\varepsilon_0 S |\vec{Q}|} e^{-\frac{Q^2 l_0^2}{2}}. \quad (2)$$

Here $a_{p,\sigma}^+$, $a_{p,\sigma}$ $b_{q,\sigma}^+$, $b_{q,\sigma}$ are the creation and annihilation operators of the electrons and of the holes correspondingly, $\varepsilon_0$ is the dielectric constant of the medium and $S$ is the surface layer area.

The wave function of two magnetoexcitons, one with quantum numbers $\vec{k}$, $\Sigma_{e1}$, $\Sigma_{h1}$ and another one with quantum numbers $-\vec{k}$, $\Sigma_{e1}$, $\Sigma_{h1}$ has the form

$$\left|\psi_{ex,ex}\left(\vec{k},\Sigma_{e1},\Sigma_{h1};-\vec{k},\Sigma_{e2},\Sigma_{h2}\right)\right\rangle = \frac{1}{N} \sum_{t,s} e^{ik_y(t-s)l_0^2} a_{t+\frac{k_x}{2},\Sigma_{e1}}^+ a_{s-\frac{k_x}{2},\Sigma_{e2}}^+ b_{-s-\frac{k_x}{2},\Sigma_{h2}}^+ b_{-t+\frac{k_x}{2},\Sigma_{h1}}^+ |0\rangle. \quad (3)$$

Here $|0\rangle$ is the ground state wave function. The wave functions of the bimagnetoexciton with the resultant wave vector $\vec{k} = 0$, or equivalently of the molecular-type bound states can be constructed as the superpositions of the wave functions (3), introducing the wave functions $\varphi_n(\vec{k})$ of the exciton relative motions in the frame of these states. The variational wave functions $\varphi_n(\vec{k})$ and their square moduli $|\varphi_n(\vec{k})|^2$ determine the distribution in the momentum space of the magnetoexcitons taking part in the formation of the bound states.

The spin of two electrons can be combined in the forms $(\uparrow\downarrow \mp \downarrow\uparrow)$ leading to the antisymmetric singlet and to symmetric triplet states correspondingly. The effective spins of two holes also can be combined in the similar structures. Below we will consider only the bound state wave functions

$$|\psi_{bimex}(0,\eta,\varphi_n)\rangle = \frac{1}{2N^{3/2}} \sum_{\Sigma_e,\Sigma_h} (\eta)^{\Sigma_e+\Sigma_h+1} \sum_{\vec{k}} \varphi_n(\vec{k}) \\ \times \sum_{s,t} e^{ik_y(t-s)l_0^2} a^+_{t+\frac{k_x}{2},\Sigma_e} a^+_{s-\frac{k_x}{2},-\Sigma_e} b^+_{-s-\frac{k_x}{2},-\Sigma_h} b^+_{-t+\frac{k_x}{2},\Sigma_h} |0\rangle, \ \eta = \pm 1, \quad (4)$$

in which the parameter $\eta = \mp 1$ corresponds to singlet-singlet and to triplet-triplet bound states respectively. They have the normalization integrals

$$\langle \Psi_{bimex}(0,\eta,\varphi_n)|\Psi_{bimex}(0,\eta,\varphi_n)\rangle = 2(1-\eta L_n(\alpha)), \ \eta = \pm 1. \quad (5)$$

The variational functions $\varphi_n(x)$ were chosen as follows

$$\varphi_0(x) = (4\alpha)^{1/2} e^{-\alpha x^2}; \ \varphi_2(x) = (8\alpha^3)^{1/2} x^2 e^{-\alpha x^2}; \ x = kl_0. \quad (6)$$

The densities of the magnetoexcitons in the frame of the bound states in the momentum space and in the real space representations are drawn in the Figure 3. The variational parameter $\alpha$ of these functions was determined so as to minimize the energy of the bound state. It is determined as the average value of the Hamiltonian (2) calculated with the wave functions (4) as follows

$$E_{bimex}(0,\eta,\varphi_n) = \frac{\langle \Psi_{bimex}(0,\eta,\varphi_n)|H^{LLL}_{Coul}|\Psi_{bimex}(0,\eta,\varphi_n)\rangle}{\langle \Psi_{bimex}(0,\eta,\varphi_n)|\Psi_{bimex}(0,\eta,\varphi_n)\rangle}. \quad (7)$$

The obtained results and the electron structure of the bound states are described below.

# 3. THE ELECTRON STRUCTURE OF THE BOUND STATES. THE RESULTS OF THE NUMERICAL CALCULATIONS

The numerical calculations with the variational wave functions (6) were effectuated. The most interesting results concern the function $\varphi_2(\vec{k})$. It describes the electron structure of the bound state, when the maximal density of the magnetoexciton cloud in the momentum space representation is concentrated on the in-plane ring with the radius $k_r = 1/(\sqrt{\alpha}l_0)$, as is represented in the Fig. 3. In the real space representation the exciton cloud has maximal density in the center of the structure with the radius of the dome $R \sim 2\sqrt{\alpha}l_0$ depending on the parameter $\alpha$.

The main formula of our paper (7) contains in the denominator the function $(1 - L_2(\alpha))$ (5), which at some values of $\alpha$ has the values lower than unity. It obliges us to calculate exactly the complicated integrals in the numerator containing one, two and three Bessel functions using the textbooks [10-12] so as to avoid the miscalculation of the fraction (7). The obtained results in the case of the variational function $\varphi_2(\vec{k})$ are drawn in the Figure 4 in dependence on the parameter $\alpha$.

They are completely different for two spin configurations with $\eta = \pm 1$. In both spin configurations the full energies of the bound states are greater than the value $-2I_l$, where $I_l$ is the ionization potential of the free 2D magnetoexciton with $\vec{k} = 0$. All these bound states are unstable as regards the dissociation in the form of two free magnetoexcitons with $\vec{k} = 0$. In spite of it, in the case $\eta = 1$ and $\alpha = 0.5$, a deep metastable bound state with considerable activation barrier comparable with the magnetoexciton ionization potentials $I_l$ was revealed. In the case $\eta = -1$ and $\alpha = 3.4$ only a shallow bound state with nonsignificant barrier appeared. All these details are reflected in Fig. 4. They were discussed initially in Ref. [9].

The results with the variational function $\varphi_0(\vec{k})$ are demonstrating very clear that the magnetoexcitons with the maximal density in the point $\vec{k} = 0$ of the momentum space practically do not interact due to the their hidden symmetry. The energies of two magnetoexcitons are very close to the value $-2I_l$ at any values of the parameter $\alpha$ with the exception of the case $\eta = 1$ and $\alpha = 0.5$, where the singularity appeared due to the value zero of the denominator.

## 4. CONCLUSIONS

The stable bound states of the two-dimensional bimagnetoexcitons in the LLLs approximation do not exist. Instead of them the metastable bound state in the triplet-triplet spin configuration with considerable activation barrier comparable with the ionization potential $I_l$ was revealed. In momentum space representation the metastable molecular state has the form with the maximal density on the ring and with the zero density in the center.

## REFERENCES

1. I.V. Lerner and Yu.E. Lozovik, J. Low Temper. Phys. **38**, 3-4, 333 (1980).
2. I.V. Lerner and Yu.E. Lozovik, Zh. Eksp. Teor. Fiz. **80**, 1488 (1981).
3. A.B. Dzyubenko and Yu.E. Lozovik, Fiz. Tverd. Tela (Leningrad) **25**, 1519 (1983); **26**, 1540 (1984) [Sov. Phys. Solid State **25**, 874 (1983); **26**, 938 (1984)].
4. E.V. Dumanov, I.V. Podlesny, S.A. Moskalenko, M.A. Liberman, Physica E **88**, 77 (2017).
5. D. Paquet, T.M. Rice and K. Ueda, Phys. Rev. B **32**, 5208 (1985).
6. S.A. Moskalenko et al., Phys. Rev. B **66**, 245316 (2002).
7. A. Wojs et al., Can. J. Phys. **83**, 1019 (2005).
8. M.A. Olivares-Robles and S.E. Ulloa, Phys. Rev B **64**, 115302 (2001).
9. S.A. Moskalenko, P.I. Khadzhi, I.V. Podlesny, E.V. Dumanov, M.A. Liberman, I.A. Zubac, Mold. J. Phys. Sci **16**, 3-4, 149 (2017).
10. G.A. Korn, T.M. Korn, *Mathematical handbook for scientist and engineers* (McGraw-Hill Book Company, New York, 1968).
11. I.S. Gradshteyn and I.M. Ryzhik. *Table of Integrals, Series, and Products* (Moscow, Nauka, 1971).
12. A.P. Prudnikov, O.I. Marichev, *Integrals and Series: Special functions* (CRC Press, 1986).
15. P.W. Anderson, Phys. Rev. **110**, 827 (1958).

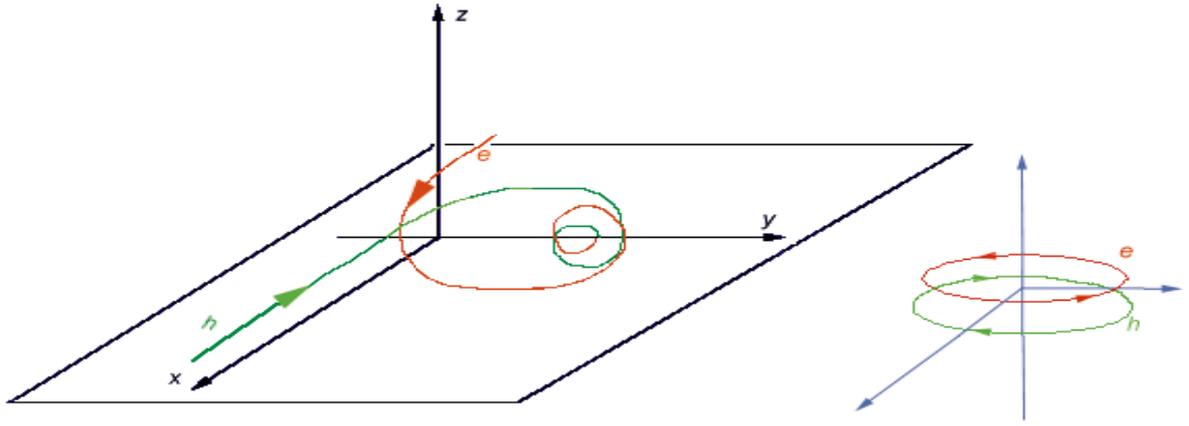

Fig. 1.

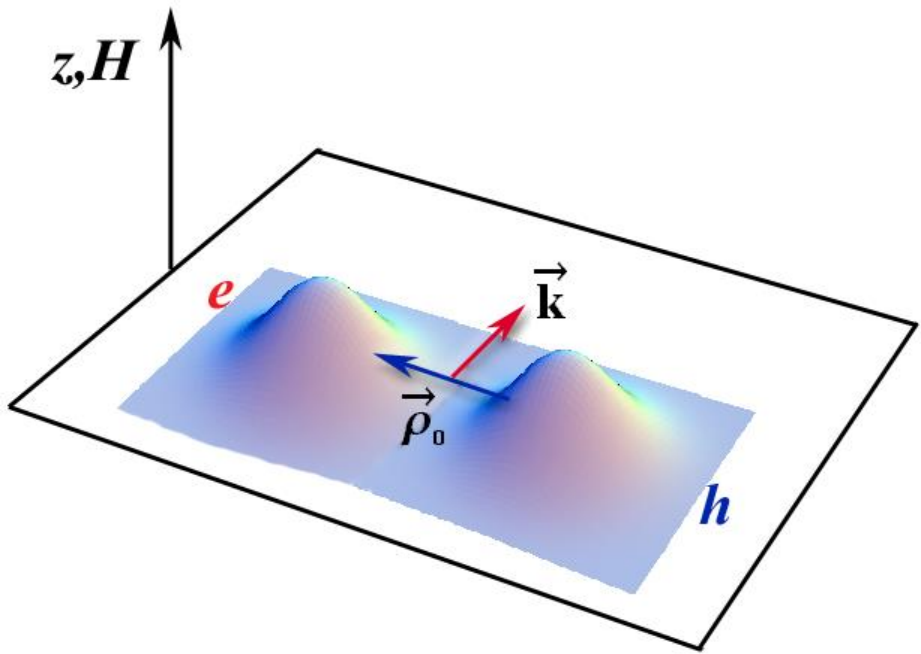

Fig. 2.

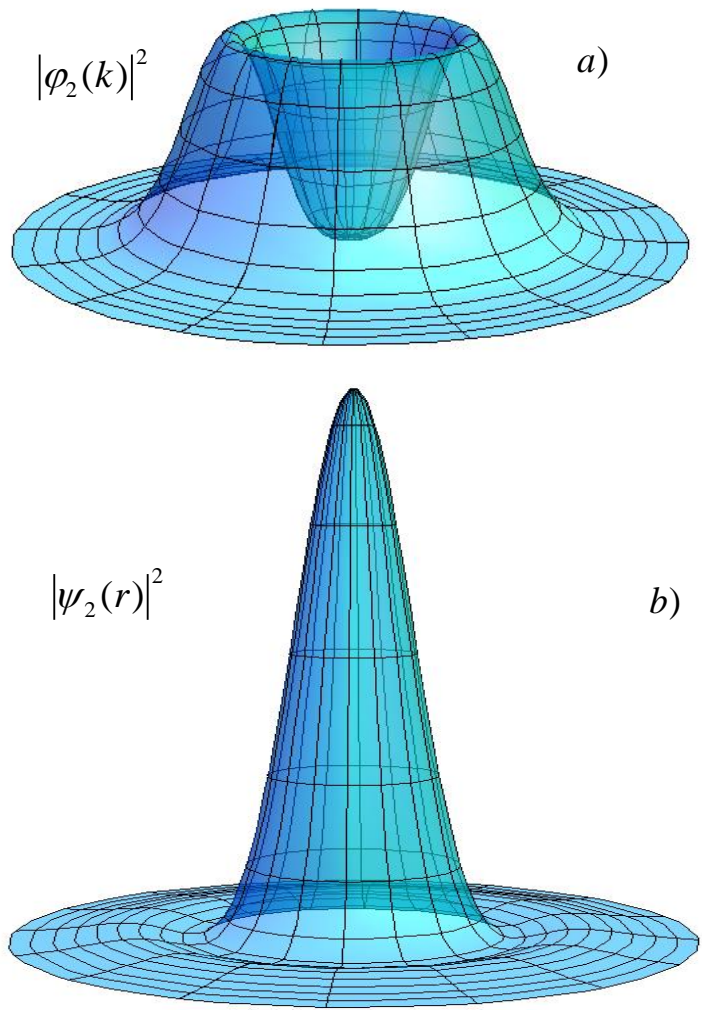

Fig. 3.

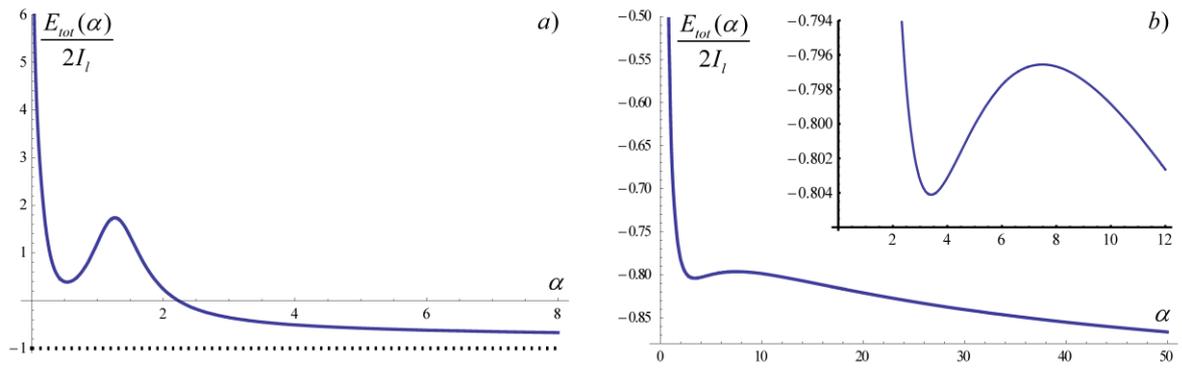

Fig. 4.

**Figure captions**

Fig. 1. The action of the Lorentz force on the in-plane moving electron-hole pair.

Fig. 2. The electric-dipole model of the 2D magnetoexciton with the wave vector $\vec{k}$ and with the arm $\vec{\rho}_0$ of the electric dipole moment.

Fig. 3. The magnetoexciton density in the frame of the bound state: a) $|\varphi_2(\vec{k})|^2$ in the momentum space representation and b) $|\Psi_2(\vec{r})|^2$ in the real space representation.

Fig. 4. The total energy of the bound states of the 2D magnetoexcitons with wave vectors $\vec{k}$ and $-\vec{k}$, with different spin structures $\eta = \pm 1$ and with the variational wave function $\varphi_2(k)$, in dependence on the parameter $\alpha$. *a)* the case $\eta = 1$ with triplet-triplet spin structure and *b)* the case $\eta = -1$ with singlet-singlet spin structure.